\documentclass[twocolumn]{aastex63}
\usepackage[caption=false]{subfig}
\turnoffedit

\submitjournal{ApJ}
\received{2021 August 30}
\revised{2021 December 7}
\revised{2022 January 7}
\accepted{2022 January 11}

\shorttitle{FRAMEx II: Simultaneous X-ray and Radio Observations of NGC~2992}
\shortauthors{Fernandez et al.}

\begin{document} 

\title{FRAMEx II: Simultaneous X-ray and Radio Variability in Active Galactic Nuclei -- \\ The Case of NGC 2992}

\correspondingauthor{Luis C. Fernandez}
\email{lfernan@gmu.edu}

\author[0000-0002-0819-3033]{Luis C. Fernandez}
\affiliation{Department of Physics and Astronomy, George Mason University, 4400 University Dr, Fairfax, VA 22030-4444, USA}
\affiliation{U.S. Naval Observatory, 3450 Massachusetts Ave NW, Washington, DC 20392-5420, USA}

\author[0000-0002-4902-8077]{Nathan J. Secrest}
\affiliation{U.S. Naval Observatory, 3450 Massachusetts Ave NW, Washington, DC 20392-5420, USA}

\author[0000-0002-4146-1618]{Megan C. Johnson}
\affiliation{U.S. Naval Observatory, 3450 Massachusetts Ave NW, Washington, DC 20392-5420, USA}

\author[0000-0003-2450-3246]{Henrique R. Schmitt}
\affiliation{Naval Research Laboratory, Remote Sensing Division, 4555 Overlook Ave SW, Washington, DC 20375, USA}

\author[0000-0002-3365-8875]{Travis C. Fischer}
\affiliation{Space Telescope Science Institute, 3700 San Martin Drive, Baltimore, MD 21218, USA}

\author[0000-0002-8736-2463]{Phillip J. Cigan}
\affiliation{Department of Physics and Astronomy, George Mason University, 4400 University Dr, Fairfax, VA 22030-4444, USA}
\affiliation{U.S. Naval Observatory, 3450 Massachusetts Ave NW, Washington, DC 20392-5420, USA}

\author[0000-0002-5604-5254]{Bryan N. Dorland}
\affiliation{U.S. Naval Observatory, 3450 Massachusetts Ave NW, Washington, DC 20392-5420, USA}

\begin{abstract}
Using simultaneous Very Long Baseline Array and Neil Gehrels Swift Observatory X-ray Telescope observations of the active galactic nucleus (AGN) in NGC~2992 over a six-month observing campaign, we observed a large drop in core 5~cm radio luminosity, by a factor of $>3$, in tandem with factor of $>5$ increase in $2-10$~keV X-ray luminosity. While NGC~2992 has long been an important object for studies of X-ray variability, our study is the first simultaneous X-ray and radio variability campaign on this object. We observe that the X-ray spectral index does not change over the course of the flare, consistent with a change in the bulk amount of Comptonizing plasma, potentially due to a magnetic reconnection event in the accretion disk. The drop in apparent radio luminosity can be explained by a change in free-free absorption, which we calculate to correspond to an ionized region with physical extent and electron density consistent with the broad line region (BLR). Our results are consistent with magnetic reconnection events in the dynamic accretion disk creating outbursts of ionizing material, increasing Compton up-scattering of UV accretion disk photons and feeding material into the BLR. These findings present an important physical picture for the dynamical relationship between X-ray and radio emission in AGNs.
\end{abstract}


\section{Introduction} \label{section:introduction}

Over the past two decades, a deep relationship between supermassive black holes (SMBHs) and their host galaxies has gradually been revealed. The discovery that the masses of SMBHs correlate tightly with the velocity dispersion of stars in the bulge of their host galaxy \citep{2000ApJ...539L...9F, 2000ApJ...539L..13G} was unexpected, as the gravitational sphere of influence of a SMBH: $r_G = G M_\mathrm{BH} \sigma_\star^{-2}$ is over two orders of magnitude too small to directly affect the dynamics of the stellar bulge. Consequently, a co-evolution between SMBHs and their host galaxies over cosmic time must occur, in which the buildup of SMBHs and (at least) classical bulges are highly correlated processes \citep{2013ARA&A..51..511K}. One of the possible mechanisms behind this co-evolution is ``feedback'' between the host bulge and SMBH, in which periods of SMBH accretion, when the SMBH radiates as an active galactic nucleus (AGN), affect the star formation efficiency of the surrounding interstellar medium (ISM). These processes invoke dynamic, causally-connected physical structures ranging from scales of $\sim$~one hundredth of a parsec out to several kpc, each emitting in some particular range of wavelengths, giving AGNs the broadest spectral energy distributions (SEDs) of any astrophysical object, effectively covering the entire electromagnetic spectrum.

In 2020, we introduced the Fundamental Reference AGN Monitoring Experiment \citep[FRAMEx;][]{2020jsrs.conf..165D}, an ongoing project led by the U.S.\ Naval Observatory to better understand the physical processes in AGNs that affect their multi-wavelength apparent positions and morphologies, such as the relationship between the accretion disk and X-ray corona, and the relationship between the X-ray corona with the production of jets and other sources of radio emission. In an initial Very Long Baseline Array (VLBA) and Neil Gehrels Swift Observatory X-ray Telescope \citep[XRT;][]{2005SSRv..120..165B} snapshot campaign of 25 AGNs that form a volume-complete sample out to 40~Mpc, we showed that the ``fundamental plane'' of black hole activity \citep[e.g.,][]{2003MNRAS.345.1057M} breaks down at high physical resolution, suggesting that core X-ray emission is (counter-intuitively) better correlated with extended radio emission and raising the prospect of truly radio-silent AGNs \citep{2021ApJ...906...88F}.


We have since followed up with a six-month monitoring campaign, in which we observed several of the AGNs detected with the VLBA on a monthly basis with simultaneous VLBA and XRT observations, in order to explore the relationship between the X-ray and radio emission at high physical resolution in the time domain. In this work, we report on our analysis of these data for NGC~2992, an Sa galaxy \citep{1991rc3..book.....D} in the early stages of a merger with its neighbor NGC~2993 \citep[e.g.,][]{2000AJ....120.1238D}. Previous work on NGC 2992 demonstrated the presence of an extended, ``Figure 8'' loops of radio emission \citep{1984ApJ...285..439U, 1988AJ.....95.1689W} symmetric about the galactic nucleus, likely a product of conical outflows \citep[e.g.,][]{1998A&A...333..459M, 2000MNRAS.314..263C, 2001AJ....121..198V} driven by the AGN \citep{2010A&A...519A..79F}, which is the dominant ionization source in the inner kpc \citep[e.g.,][]{2021MNRAS.502.3618G}. 

NGC~2992 is particularly variable in the X-rays, exhibiting $2-10$~keV luminosity changes by over an order of magnitude \citep[e.g.,][]{2000A&A...355..485G}, sometimes within days-to-weeks timeframes \citep{2007ApJ...666...96M}. This extreme variability has been implicated in the spectral changes that NGC~2992 exhibits at visual wavelengths, varying between a Seyfert 1.5 and and a Seyfert~2 without measurable changes in line-of-sight reddening \citep{2008AJ....135.2048T}, tying the $2-10$~keV emission to ionizing continuum variability. At hard X-ray energies ($>10$~keV), NGC~2992 presents a simple power-law spectrum that exhibits variability similarly to the soft X-ray emission \citep{2007ApJ...666..122B}. When considered jointly with the soft X-ray emission, both the power-law spectral index $\Gamma$ and the hydrogen column density $N_H$ are relatively constant over time, leading \citet{2007ApJ...666..122B} to argue that the X-ray luminosity variability in NGC~2992 could be due to varying amounts of coronal plasma, as might be provided by flares attributed to magnetic reconnection events in the SMBH accretion disk.

X-ray emission line spectroscopy has shed light on the nature of the X-ray flares in NGC~2992, with \citet{2007ApJ...666...96M} noting that a highly redshifted, broad Fe~K$\alpha$ line appears during periods of high X-ray luminosity, as later confirmed by \citet{2010ApJ...713.1256S}, in contrast with Fe~K$\alpha$ emission at 6.4~keV normally found in the X-ray spectrum and attributed to more distant matter, as noted also in \citet{2007PASJ...59S.283Y}. Other K-shell line emission lines of Si and S near their rest-frame energies were later detected using higher spectral resolution observations during a period of lower X-ray luminosity, but these observations also demonstrated the existence of Si lines redshifted by 2500~km~s$^{-1}$, indicating powerful AGN-driven outflows in NGC~2992 \citep{2017ApJ...840..120M}. In X-ray absorption, \citet{2018MNRAS.478.5638M} recently found tentative evidence for an ``ultra fast outflow'', with material being ejected from the innermost accretion regions with a velocity of about $0.21c$, and a triggered simultaneous broadband campaign by \citet{2020MNRAS.496.3412M} also showed a transient emission line at 5.4~keV, which the authors attributed to a component just a few gravitational radii away from the black hole in the highly variable accretion disk.

The overall picture of NGC~2992 then is one of a powerful AGN, possibly triggered by the merger with NGC~2993, that is driving material out of the center of the galaxy to large physical scales. The AGN is highly variable, with several distinct kinematic components, changes in spectral type (between type 1 and type 2) over short timescales, providing a direct probe into the innermost accretion processes of AGNs. The AGN in NGC~2992 is therefore an excellent case study for models of AGN accretion, as was noted by \citet{2007IAUS..238..123Y} and \citet{2007ApJ...666...96M}. 

Despite the rich multi-wavelength variability demonstrated in this AGN, to date no study of the radio variability in NGC~2992 has been conducted. This is a major deficiency, as while the X-ray emission in AGNs is a proxy for the thermal accretion luminosity (via inverse Compton scattering of disk UV photons), the radio emission traces the non-thermal, magneto-hydrodynamic interaction of BH/accretion disk magnetic fields and high-energy electrons responsible for putative jet activity \citep[for a review, see][]{2019ARA&A..57..467B}. Moreover, while there have been entire VLBI monitoring campaigns of luminous AGN jets \citep[e.g.,][]{2009AJ....137.3718L,2016AJ....152...12L}, there have been only a handful of VLBI monitoring campaigns of nearby radio-quiet AGNs \citep[e.g.,][]{2003ApJ...591L.103B,2021MNRAS.504.3823W}, and to our knowledge none with simultaneous X-ray and radio monitoring.


In this paper, we discuss the results of a time-domain campaign to study the simultaneous VLBA C-band (5~cm) core radio and Swift~XRT $0.2-10$~keV X-ray properties of the AGN in NGC~2992. The main goal of this paper is to explore the temporal relationship between radio core and X-ray emission in AGNs, and what this relationship reveals about the accretion process in AGNs. In Section~\ref{section: methods}, we describe our observation campaign and the VLBA and Swift XRT data analysis. We present our results in Section~\ref{section: results}, provide discussion in Section~\ref{section: discussion}, and give our main conclusions in Section~\ref{section: conclusions}.


\section{Methodology} \label{section: methods}
We use a redshift of $z= 0.00771$ \citep{1996ApJS..106...27K} and a distance of $D = 33.2$~Mpc for NGC~2992, as in \citet{2021ApJ...906...88F}, giving an angular scale of 0.16~pc~mas$^{-1}$. For the black hole mass, we use the calcium triplet-based measurement of $\sigma_\star=154$~km~s$^{-1}$ from \citet{2020A&A...634A.114C} and the $M_\mathrm{BH}-\sigma_\star$ relation from \citet{2013ARA&A..51..511K}, which gives a logarithmic $M_\mathrm{BH}\sim8.00$, with an intrinsic scatter-based uncertainty of $0.28$~dex.\footnote{We note that this black hole mass is 2.2 times lower than the value adopted in \citet{2021ApJ...906...88F}.} The corresponding Eddington luminosity is $1.3\times10^{46}$~erg~s$^{-1}$. The bolometric luminosity of the AGN in NGC~2992 was estimated at $8.3\times10^{43}$~erg~s$^{-1}$ by \citet{2002ApJ...579..530W}, who compiled a catalog of black hole masses and bolometric luminosities for AGNs. A comparison with similarly-derived bolometric luminosities from \citet{1988A&A...205...53P} indicates a dispersion of $\sim0.33$~dex \citep[Figure~5 in][]{2002ApJ...579..530W}, and the value from \citet{2002ApJ...579..530W} is consistent with that derived in \citet{2015MNRAS.449.1309G}. The Eddington ratio of the SMBH in NGC~2992 is therefore $L_\mathrm{bol}/L_\mathrm{Edd}\sim0.0064$, with an uncertainty of about 0.44~dex.

\subsection{\textit{Very Long Baseline Array} Observations} \label{subsection: VLBA}

We received observation time through the U.S.\ Naval Observatory's 50\% timeshare (PI:~T.\ Fischer) allocation on the VLBA telescope at 5 cm (6 GHz) every 28 days starting December 31, 2019 to perform a total of 6 observations at regular intervals. Utilizing the same phase referencing method from the initial FRAMEx snapshot survey \citep{2021ApJ...906...88F}, we altered telescope pointings between our target and a nearby known phase reference calibrator.  In this way, we are able to accurately constrain the position and phase of NGC 2992.  We move between our target and the phase calibrator after 4 and 2 minute integration times, respectively. We observed NGC 2992, along with two other targets of similar right ascension from our FRAMEx series, which are also observed using the phase referencing technique.  We cycled between the three targets in 20 minute on-source time intervals in order to maximize $uv$ coverage so that we can produce high fidelity images for each source.  We observed NGC~2992 for a total integration time of 1 hour and requested to use all 10 VLBA antennas in each observing session. Unfortunately, the two observing sessions on 12/31/2019 and 04/21/2020 did not use all 10 antennas; 12/31/2019 did not use HN and OV and 04/21/2020 did not use MK. The results of the other two FRAMEx targets observed along with NGC 2992 will be presented in a future work. 

We used the new, shared-risk, Mark 6 data recorders, which provide 4 Gbps recording rate and enable dual polarization observations over 512 MHz of continuous bandwidth.  We used 2-bit sampling and four, 128 MHz, intermediate frequency (IF) windows each with 512 channels for a spectral resolution of 250 kHz. Due to the shared-risk nature of the Mark 6 data recorders, during our February and March 2020 set of observations, there was an unforeseen software bug in the NRAO's configuration of the recorders, which affected our 6 GHz setup (Section~\ref{subsubsection: Analysis}).  The results of this software bug are that we lost the fourth IF window for observations during these two months and there was a residual amplitude scaling issue that affected our flux measurements.  We carefully addressed these issues in Section \ref{subsubsection: Analysis}.  Table \ref{tab:radio_obs} presents our observation parameters for the six sessions we observed NGC 2992.

\begin{deluxetable*}{lclccccccc} \label{tab:radio_obs}

\tablehead{ & \colhead{Antennas}  &\colhead{T$_{\rm int}$} & \colhead{F$_{\rm center}$} &               \colhead{Bandwidth} & \colhead{F$_{\rm range}$} & \colhead{Restoring Beam} &                \colhead{Beam angle} & \colhead{RMS} & \colhead{RMS$_\mathrm{theoretical}$}\\  [-0.2cm]
            \colhead{Date} & \colhead{(Missing)} &\colhead{(s)} &\colhead{(GHz)} & \colhead{(MHz)} & \colhead{(GHz)} & \colhead{($\alpha \times \delta$; mas)}	& \colhead{(deg)} & \colhead{($\mu$Jy bm$^{-1}$)} & \colhead{($\mu$Jy bm$^{-1}$)} }

\startdata
31 Dec 2019 & HN,OV & 2564 & 5.803879 & 384 & 5.612$-$5.996 & 7.23$\times$3.32 & -8.0 & 45 & 29 \\
27 Jan 2020 & None  & 3268 & 5.805074 & 384 & 5.612$-$5.996 & 7.56$\times$2.64 & ~4.2 & 42 & 20 \\
25 Feb 2020 & None  & 3036 & 5.801782 & 384 & 5.612$-$5.996 & 8.02$\times$2.16 & -6.7 & 56 & 21 \\
25 Mar 2020 & None  & 2796 & 5.796680 & 384 & 5.612$-$5.996 & 9.80$\times$4.33 & ~2.9 & 57 & 22 \\
21 Apr 2020 & MK    & 2788 & 5.808148 & 384 & 5.612$-$5.996 & 7.76$\times$3.21 & -5.7 & 52 & 24 \\
19 May 2020 & None  & 3260 & 5.806133 & 384 & 5.612$-$5.996 & 8.05$\times$2.99 & -3.8 & 45 & 20 \\
\enddata

\caption{VLBA Observations. Target: NGC 2992; Phase calibrator: J0941$-$1335.}

\end{deluxetable*}

\begin{deluxetable*}{cccccc}  \label{tab:vlbadata}

\tablehead{\colhead{Observation Date}	& \colhead{$I_\mathrm{\nu ~peak}$}  &                                \colhead{$F_\mathrm{peak}$} & \colhead{log$_{10}(L_\mathrm{peak}\  / \mathrm{\ erg} \mathrm{\ s}{^{-1}})$} &                            \colhead{$S_\mathrm{int}$}  & Noise RMS\\  [-0.2cm] &
           \colhead{(mJy~beam{$^{-1}$})}	& \colhead{({${\times}10^{-16}$} erg~s{$^{-1}$} cm{$^{-2}$})} 	&  &   \colhead{(mJy)} & \colhead{(mJy)}
            }

\startdata
12/31/19             & 1.00$\pm$0.09  & 0.58$\pm$0.05 & 36.88      & 1.5$\pm$0.2  & 0.045\\
01/28/20             & 0.93$\pm$0.08  & 0.54$\pm$0.05 & 36.85      & 1.8$\pm$0.2  & 0.042\\
02/25/20{$^a$}     & 0.3$^{+0.3}_{-0.2}$ & 0.2$^{+0.2}_{-0.1}$ & 36.40 & 0.3$^{+0.4}_{-0.2}$ & 0.056\\
03/25/20{$^a$}     & $<$ 0.5{$^b$}    & \nodata         & \nodata    & \nodata          & 0.057\\
04/21/20             & 1.2$\pm$0.1  & 0.68$\pm$0.06 & 36.95      & 2.0$\pm$0.2  & 0.052\\
05/19/20             & 0.96$\pm$0.09  & 0.56$\pm$0.05 & 36.86      & 1.3$\pm$0.2  & 0.045\\
\textbf{C} & 1.01$\pm$0.05  & 0.58$\pm$0.03 & 36.89      & 1.62$\pm$0.06  & 0.026\\
\enddata

\noindent Peak intensity and integrated flux density calculated using the AIPS task {\sc jmfit} which uses an elliptical Gaussian fitting algorithm. All uncertainties listed are $\pm2\sigma$.\\
$^a$Corrected using scaling factors where uncertainties are calculated using Monte Carlo to obtain 95.4\% confidence interval.\\ 
$^b$Upper limit shown using 3$\sigma$ over the RMS.

\caption{\centering{6 GHz VLBA Measurements of NGC 2992. Row \textbf{C}: Epochs unaffected by software bug concatenated together and imaged to obtain NGC 2992's 6-month average peak/integrated flux density and luminosity. For the observation on 02/25/20, this is a marginal detection since it does not hold at a 3$\sigma$ level. Values for this epoch are shown to distinguish it from the completely non-detected epoch. }}

\end{deluxetable*}

\subsubsection{Calibration}

To calibrate the VLBA data obtained from our observations, we used NRAO's software package, Astronomical Image Processing System (AIPS) \citep{2003ASSL..285..109G} release \texttt{31DEC19}. We first loaded in the data with a calibration (CL) table interval of 0.1 minutes. Next, we used the {\sc vlbautil} module which corrected for the ionospheric delays and Earth orientation parameters. Before the sample threshold errors were corrected, we used the task {\sc tysmo} to clip the system temperature (T$_{\rm sys}$) table (TY) values above a factor of $\sim2$ times the average T$_{\rm sys}$ values over the duration of the observation. We flagged T$_{\rm sys}$ values below 0~K since these are either non-physical or from other instrumental effects. This task then replaced the clipped values by interpolating across the clipped region using a linear interpolation function. In addition to flagging spurious system temperatures, we also flagged data below $15\arcdeg$ elevation. Next, we \edit1{flagged} out any high amplitude RFI as a function of time using the task {\sc editr} and then \edit1{flagged} RFI as a function of frequency using the task {\sc wiper}. We note that the Brewster and Kitt Peak antennas consistently had major RFI throughout all observations, which required significant flagging. Once flagging was complete, we calibrated for correlator sampler threshold errors, instrument delays, bandpass, amplitude, and parallactic angle. We checked to see if there were any bad solutions applied to the CL table using {\sc editr} and flagged them out. Next, we solved for phase and complex amplitudes with the task {\sc fring} and applied the solutions to the CL table for both the phase calibrator and source using the task {\sc clcal}. We checked the last calibration table using {\sc editr} and {\sc wiper} again and flag out any bad solutions. Finally, we apply the phase calibrator's CL table to the source using the task {\sc split} and a two-point interpolation function, thus preserving the calibrated phase and absolute astrometry to the accuracy level of the phase calibrator's position. We used the phase calibrator J0941-1335 from the ICRF3 catalog \citep{2020A&A...644A.159C}, at position $\alpha=145\fdg2606228293$, $\delta=-13\fdg597495756639$. Before imaging, we flagged any remaining high amplitudes on the source and calibrator using the task {\sc wiper}.

\subsubsection{Imaging}

\begin{figure*}
  \includegraphics[width=\textwidth]{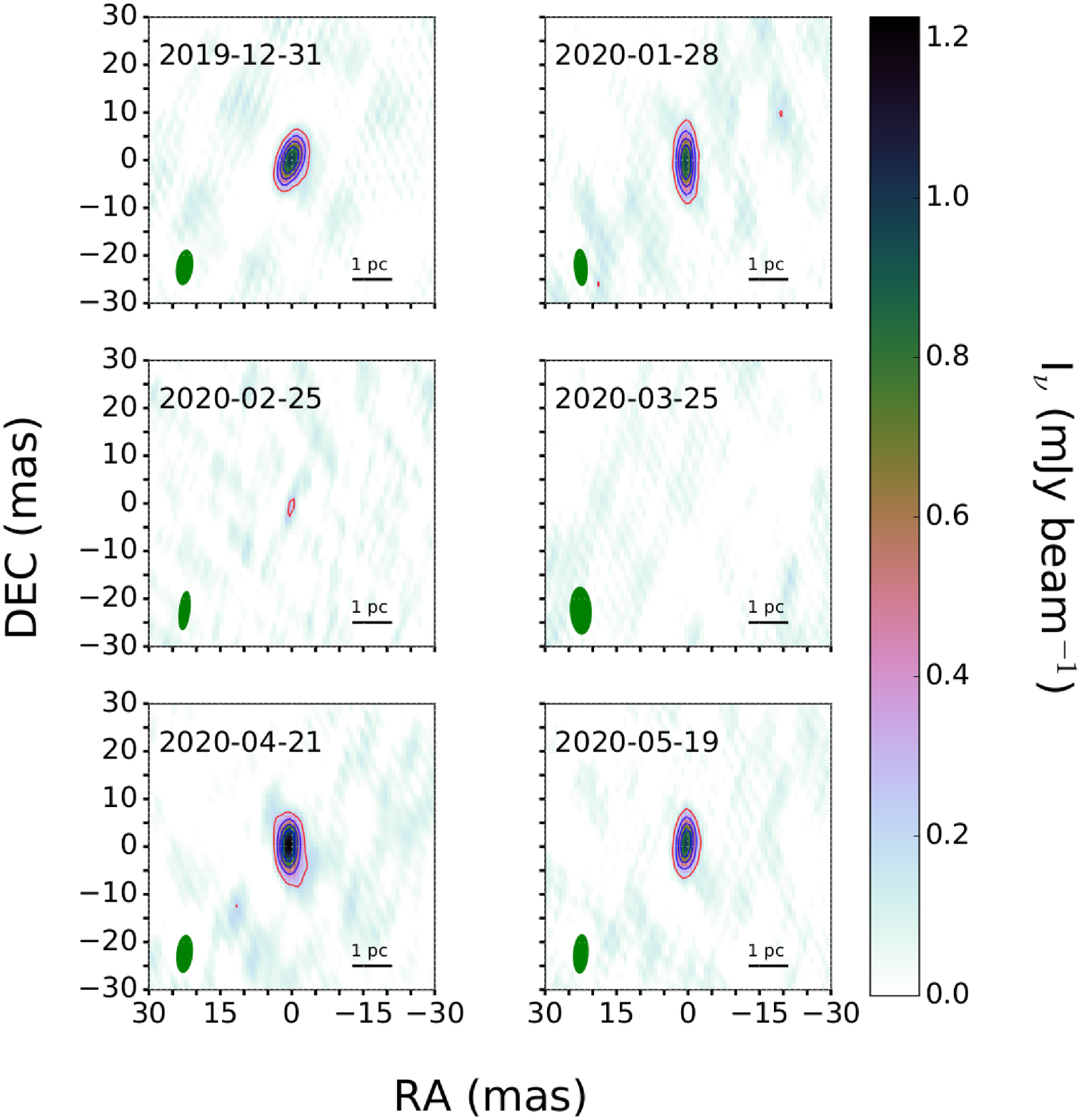} 
  \caption{6 GHz (5 cm) images of NGC 2992. Position centered at RA(J2000) = 146.42476 and DEC(J2000) = -14.326274. Red contour is \edit1{4 times the image RMS} with subsequent contours at 8, 12, and  \edit1{16 times image RMS} in blue. The restoring beam is to the lower left of each image in green. (For the observations on 2020-02-25 and 2020-03-25, these images have been corrected using the scaling factor.)}
  \label{fig: Imaging}
\end{figure*}

To image the calibrated data, we use the AIPS task {\sc imagr}. We set the cell size to 0.8~mas and the image size to $512\times512$ pixels. This makes the field of view for each image $0\farcs41\times0\farcs41$. Next, we set the Briggs robustness close to natural weighting to obtain the highest sensitivity, in order to determine if there is any variability in structure and flux. When imaging, we interactively set a box around the source and perform iterations of deconvolution of the point spread function (the CLEAN algorithm), until the root mean square (RMS) of the residuals inside the box are the same as the RMS outside the box. Self-calibration is typically used to improve the quality of an image with high signal-to-noise (S/N). This involves using the task {\sc calib} to first improve phase and the RMS thermal noise. Then, repeat this process until the S/N does not improve any further. Finally, use the task again to improve amplitude and phase to make the final self-calibrated and cleaned image. We applied self-calibration to all of our phase calibrators and to the other two FRAMEx targets observed during our 6 sessions.  Unfortunately, self-calibration was not possible for NGC~2992 because of its low S/N (Figure \ref{fig: Imaging}). 

\begin{figure*}
  \hspace{-0.5cm}
  \includegraphics[width=\textwidth]{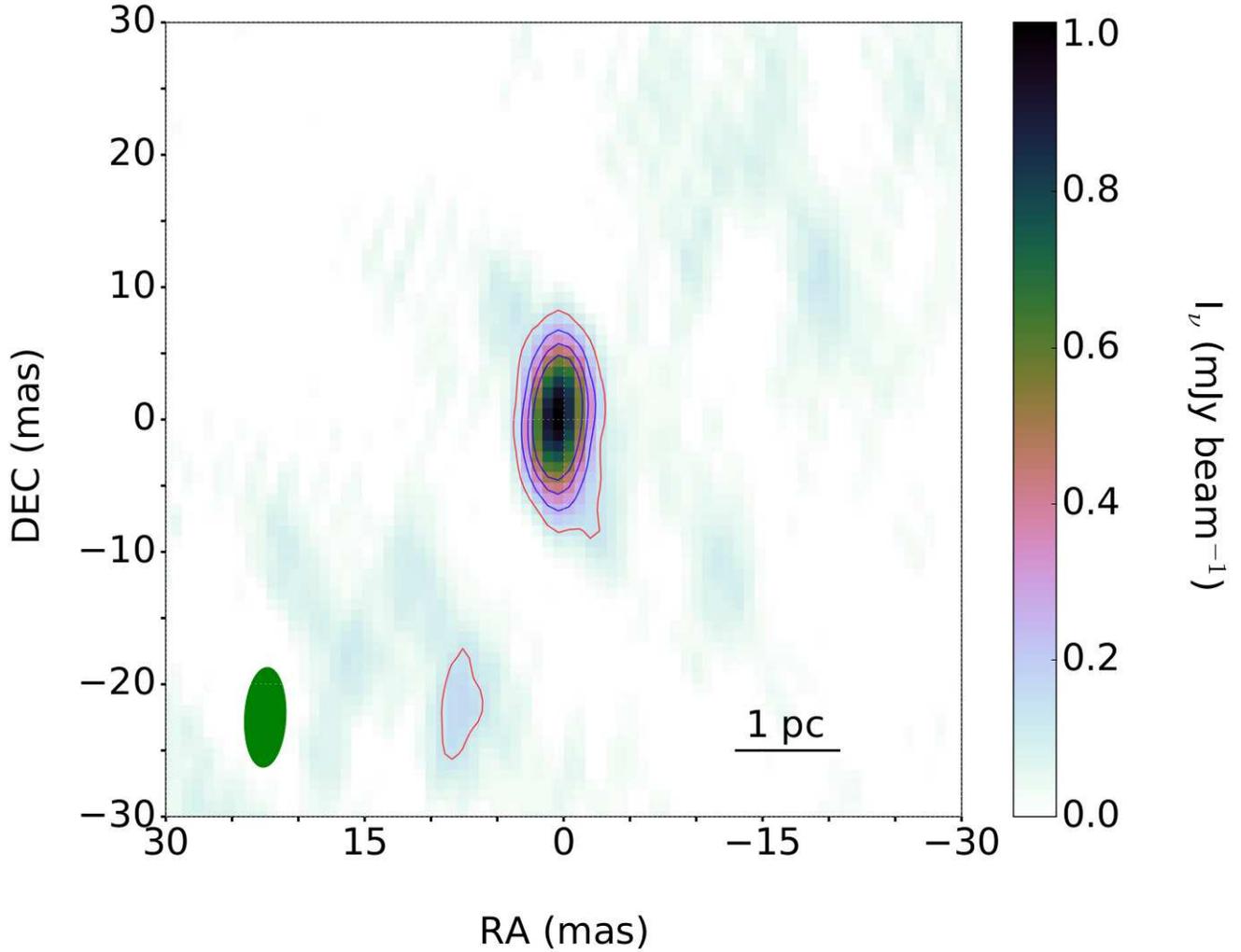}
  \caption{6 GHz concatenated image of NGC~2992 using only the non-affected data. Position centered at RA(J2000)~$= 146.42476$ and DEC(J2000)~$= -14.326274$. Red contour is \edit1{4 times the image RMS} with subsequent contours at 8, 12, and \edit1{16 times image RMS} in blue. The restoring beam is to the lower left in green. There is a faint radio emission to the SE of the primary radio source with a significance of \edit1{$\geq 5$ times the image RMS.} It is unclear as to how extended this emission is.}
  \label{fig: Imaging Avg}
\end{figure*}

To examine the source structure in detail and maximize sensitivity for our observations, we concatenated the data unaffected by the software bug (see Section~\ref{subsubsection: Analysis}) and the resulting image is shown in Figure \ref{fig: Imaging Avg}.

\begin{deluxetable*}{crrcccc}
\tablehead{\colhead{ID} & \colhead{R.A.} & \colhead{Decl.} & \colhead{Mode} & 
           \colhead{UTC Start} & \colhead{UTC Stop} & \colhead{Obs. Time} \\ [-0.2cm]&  & \colhead{deg}  & \colhead{deg}   & &  & \colhead{second}
           }

\startdata
35344 & 146.4247756 & -14.32626689 & PC & 12/31/19 12:54 & 12/31/19 15:25 & 1692\\
35344 & 146.4247756 & -14.32626689 & PC & 01/28/20 10:13 & 01/28/20 11:08 & 1682\\
35344 & 146.4247756 & -14.32626689 & PC & 03/25/20 09:38 & 03/25/20 10:33 & 1687\\
35344 & 146.4247756 & -14.32626689 & PC & 05/19/20 02:30 & 05/19/20 03:25 & 1707\\
\enddata

\caption{{\it Swift} XRT Observations of NGC 2992}
\end{deluxetable*}

\subsubsection{Radio Analysis} \label{subsubsection: Analysis}

Using NRAO's AIPS software, we used the task {\sc jmfit} to analyze each epoch's final cleaned image. \edit1{After setting the parameters to search inside a designated box, this task used an elliptical Gaussian fitting algorithm with the image's RMS to calculate the FWHM of the source to obtain the peak and integrated flux densities with their respective 1$\sigma$ uncertainty. To account for any systematic uncertainties and for a conservative treatment of the total uncertainty in these data, we used two times the 1$\sigma$ errors and denote it as 2$\sigma$. Table \ref{tab:vlbadata} contains the results for the VLBA measurements.}



Systematic issues came to light during our analysis for two data sets obtained in February and March in comparison to the other observations. Specifically, the source amplitudes and RMS thermal noise values are systematically lower for these months across all of our targets observed in the sessions containing NGC 2992. As was previously noted in section \ref{subsection: VLBA}, a software bug impacted our observations.  We used new Mark 6 data recorders under ``shared-risk,'' thus assuming responsibility for any systematics in the new recording system that might cause issues. Through private communications with several NRAO experts responsible for managing the VLBA, we learned that the software bug was thought to have been associated with repeated drop-outs in the data recorders.  
However, if the recording drop-out was the only issue, then the amplitudes for each of the calibrators and targets should have been corrected when self-calibration was applied.  Unfortunately, self-calibration did not improve the amplitudes nor RMS thermal noise properties in any of the phase calibrators.  Thus, there were likely issues beyond those identified by NRAO. The instrumental effects went beyond coherence losses as these would have been corrected during self-calibration and we would have recovered the expected amplitudes for each of our targets. Because our data continued to suffer from these dramatic intensity offsets during the months of February and March, we turned to an analytic approach to correcting the data.

\begin{figure}
    \includegraphics[width=\columnwidth]{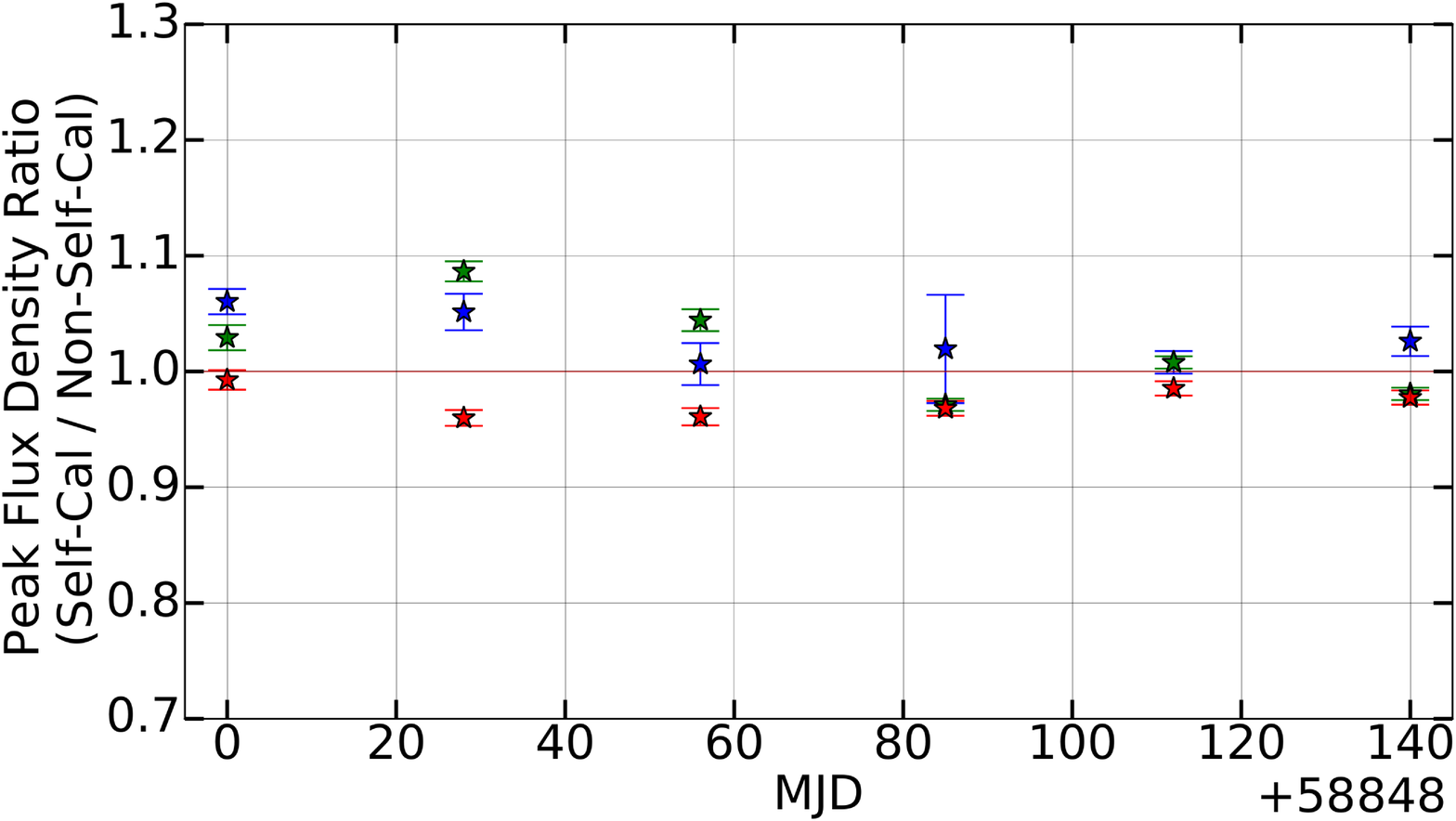}
    \caption{Phase Calibrator ratios of peak flux density from self-calibrated and non-self-calibrated images. Red stars are for J1206+3941, green stars are for J0956+5753, and blue stars are for J0941-1335.}
  \label{fig: Peak Ratio}
\end{figure}

We examined the peak intensities for all objects observed during the same observations as NGC 2992. These include the following phase calibrators and respective targets: PMN J0941-1335 for NGC 2992, JVAS J0956+5753 for NGC 3079, and JVAS J1206+3941 for NGC 4151. Figure \ref{fig: Peak Ratio} shows the ratio between self-calibrated and non-self-calibrated peak intensities for the three phase calibrators.  Since these ratios are consistent with one, we use the non-self-calibrated peak intensities for all sources in our analysis as we are unable to self-calibrate the observations of NGC~2992 due to its low intensity. \edit1{Because the ratios in Figure \ref{fig: Peak Ratio} are consistent with $\sim$1 across all epochs, this implies that self-calibration is inadequate to correct for the amplitude drops for the problematic sessions in the months of February and March.}

\begin{figure}[h]
    \includegraphics[width=\columnwidth]{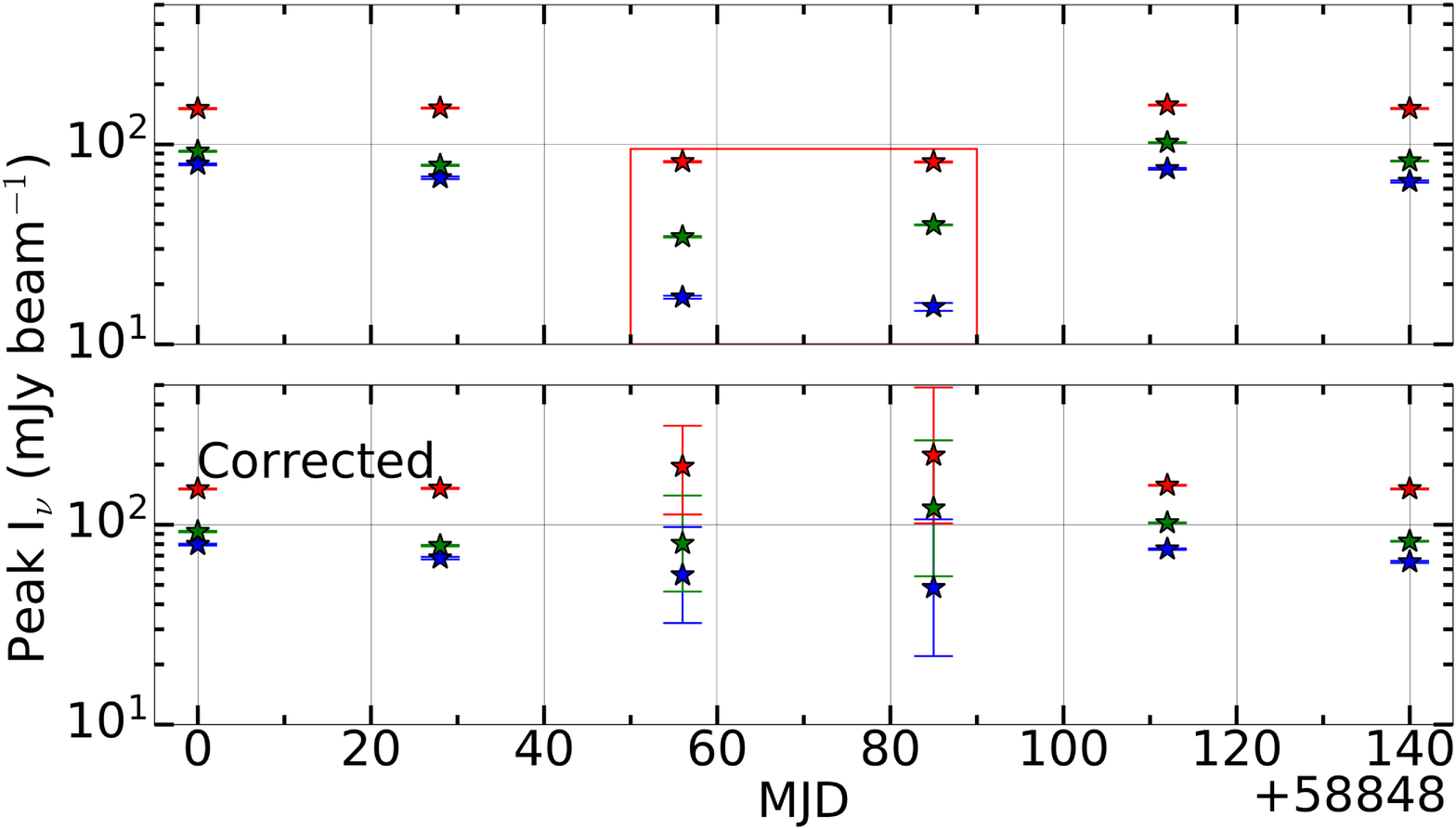}
    \caption{Top: Peak intensities for phase calibrators (all uncertainties are $\pm2\sigma$). Red points are for J1206+3941. Green points are for J0956+5753. Blue points are for J0941-1335. The enclosed red box delineates sessions affected by the spectral window 4 software bug (Section~\ref{subsubsection: Analysis}). Bottom: Corrected with their respective scaling factors. Uncertainties for affected observations calculated using Monte Carlo to obtain 95.4\%confidence intervals.}
  \label{fig: Phase_Cal}
\end{figure}

The top of Figure \ref{fig: Phase_Cal} shows the peak intensities of the phase calibrators for all observations including the problematic sessions in February and March. We find that every object observed during these months exhibit a similar decrease in peak intensity. As it is unknown if both observations are affected equally, we need to determine a scaling factor for each one. In order to ascertain whether we are obtaining the correct scaling factor, we devised two independent methods to estimate it.

The first method utilizes the peak intensities $I_\nu$ to calculate the ratio of the non-affected data from that of the affected data for each phase calibrator and source (excluding NGC~2992). We take the ratio of intensities for each affected observation $a$ to the non-affected across observations, giving a scale factor $f_{\rm peak} = {I_\nu} / I_{\nu,a}$, where we use the inverse variances $\sigma_{I_{\nu,a}}^{-2}$ as weights. Using standard error propagation, the uncertainty of the scaling factor is:

\begin{equation} \label{eq: Scaling Factor}
\sigma_{f_{\rm peak}}^2 = \left(\frac{\sigma_{{I_\nu}}}{I_{\nu,a}}\right)^2 + \left(\frac{{I_\nu}~\sigma_{I_{\nu,a}}}{I_{\nu,a}^2}\right)^2
\end{equation}

\noindent where $I_{\nu,a}$ is the peak intensity for the affected data and $\sigma_{I_{\nu,a}}$ is its associated uncertainty. This allowed us to obtain \edit1{a total of 8 scaling factors for each object based on comparing the 4 non-affected observations with the 2 affected observations (4 scaling factors for each affected epoch).}

The second method examines a cleaned, non-self-calibrated image of every calibrator and source (excluding NGC 2992) to compare the RMS from the noise of the non-affected with the affected observations. 
The first issue that needs to be corrected for before comparing the RMS values is the fact that not all observations utilize all 10 VLBA antennae. To account for this, we take the ratio of the observed RMS to the theoretical RMS for each observation, calculated using the number of antennae used, integration times, and the total used bandwidth. This enables us to have the most accurate comparison between all observations. Our final scaling factor, $f_{\rm {rms}}$, for the RMS method is derived using the following:

\begin{equation}
    {f_{\rm {rms}}} = \frac{(\rm{RMS}_{\rm meas}/\rm{RMS}_{\rm theo})_\nu}{(\rm{RMS}_{\rm meas}/\rm{RMS}_{\rm theo})_{\nu,a}}
\end{equation}

\begin{figure}[h]
    \includegraphics[width=\columnwidth]{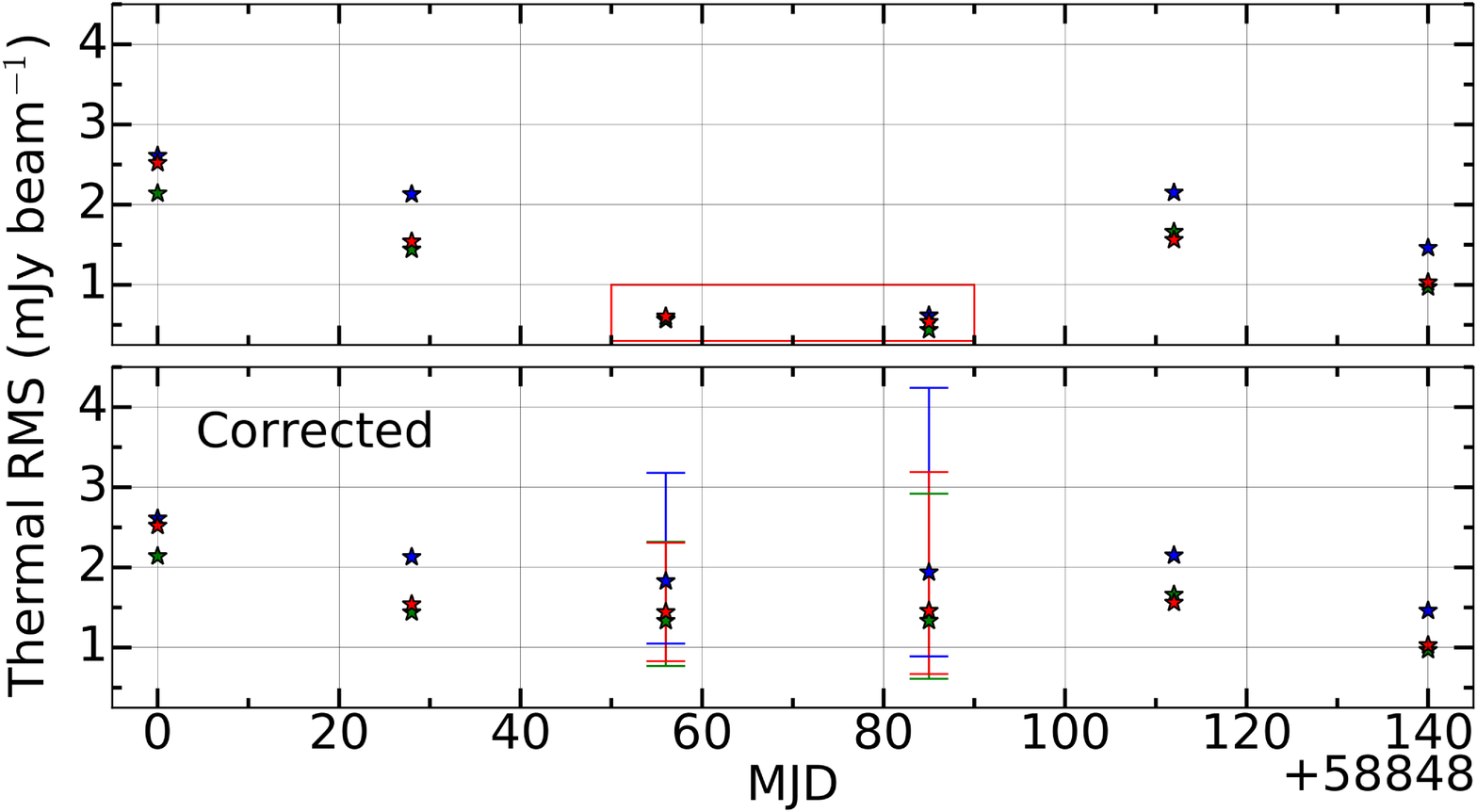}
    \caption{Phase Calibrator's thermal RMS from images using a small number of iterations in CLEAN. Top: Before Corrections. Bottom: After corrections using their respective scaling factors. Uncertainties for affected observations calculated using Monte Carlo to obtain 95.4\% confidence intervals. Red stars are for J1206+3941, green stars are for J0956+5753, and blue stars are for J0941-1335. Enclosed red box: Delineation caused by systematic issues.}
  \label{fig: DB Phase Cal}
\end{figure}

\begin{figure}[h]
    \includegraphics[width=\columnwidth]{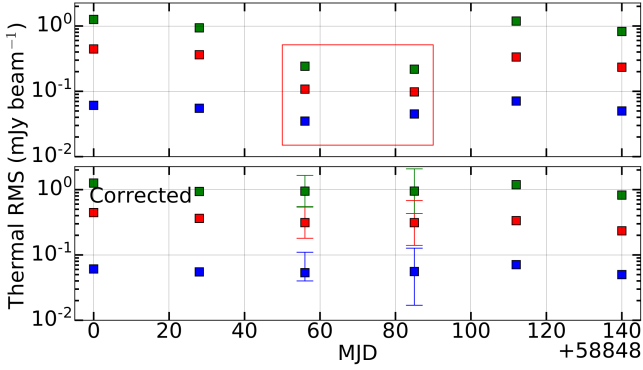}
    
    \caption{Source's thermal RMS from images using a small number of iterations in CLEAN. Top: Before Corrections. Bottom: After corrections using their respective scaling factors. Uncertainties for affected observations calculated using Monte Carlo to obtain 95.4\% confidence intervals. Blue squares are for NGC 2992, Red squares are for NGC 4151, and green squares are for NGC 3079. Enclosed red box: Delineation caused by systematic issues. }
  \label{fig: DB Sources}
\end{figure}

We then find the individual scaling factors from each non-affected observation to be consistent with the first method where we obtain 4 scaling factors for each \edit1{object's affected epoch.} Since the associated uncertainties of the RMS ratios are negligible, variance in the scaling factor from the RMS method represents real dispersion. Figure \ref{fig: DB Phase Cal} shows the uncorrected (top panel) and corrected (bottom panel) RMS thermal noise values for all six epochs for each phase calibrator.  Similarly, Figure \ref{fig: DB Sources} shows the uncorrected (top panel) and corrected (bottom panel) RMS thermal noise for the other two FRAMEx sources, NGC 3079 and NGC 4151, and NGC 2992.

\begin{figure}[h]
    \includegraphics[width=\columnwidth]{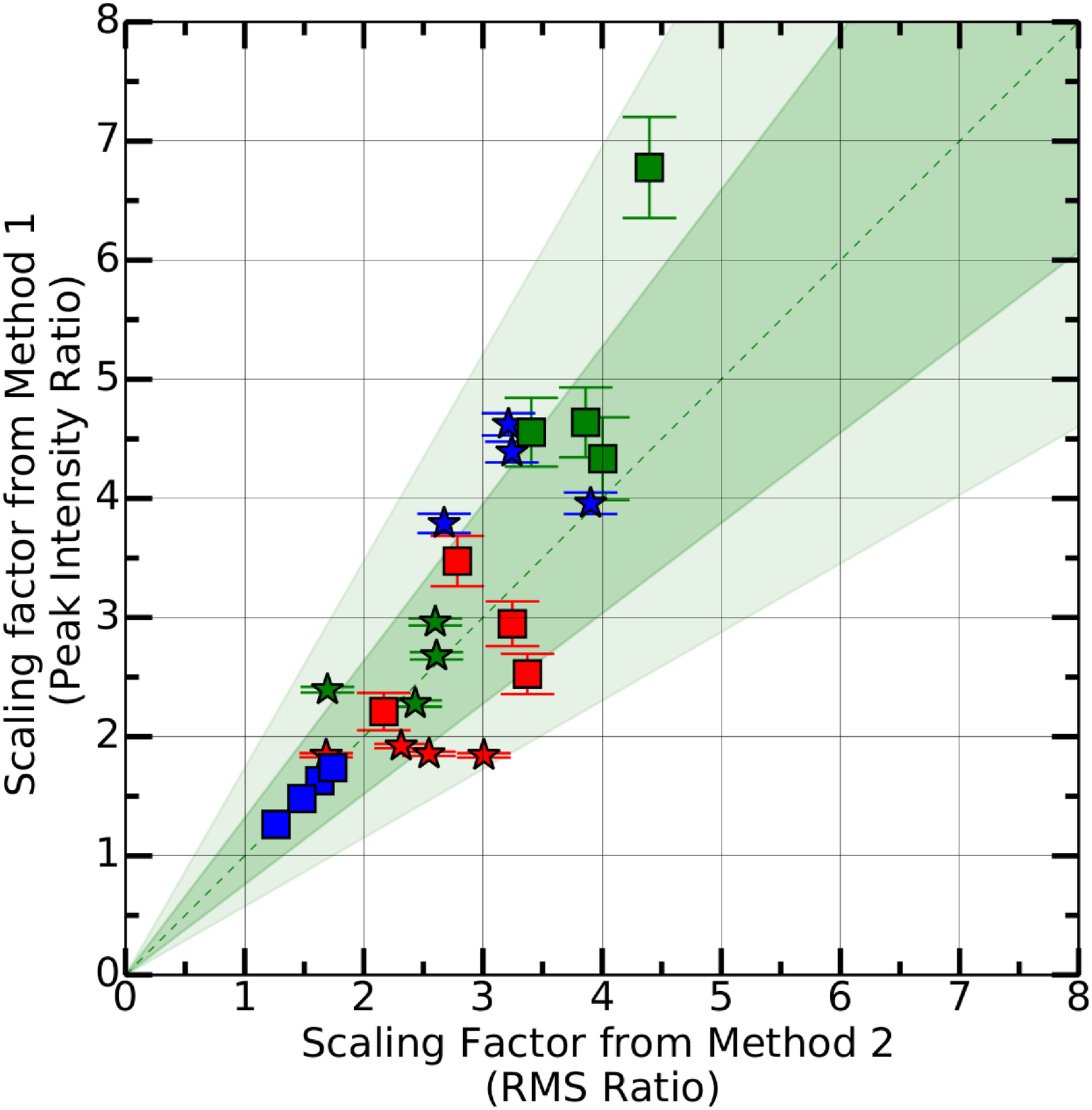}
    
    \caption{Scaling factors from Method 1 vs.\ Method 2 for observations on 02/25/2020. Darker shaded region is the $1\sigma$ dispersion (0.12~dex) while the lighter shaded region is the $2\sigma$ dispersion. Red stars are for J1206+3941. Green stars are for J0956+5753. Blue stars are for J0941-1335. Blue squares are for NGC 2992, red squares are for NGC 4151, and green squares are for NGC 3079. 
    }
  \label{fig: Method Ratio C}
\end{figure}

\begin{figure}[h]
    \includegraphics[width=\columnwidth]{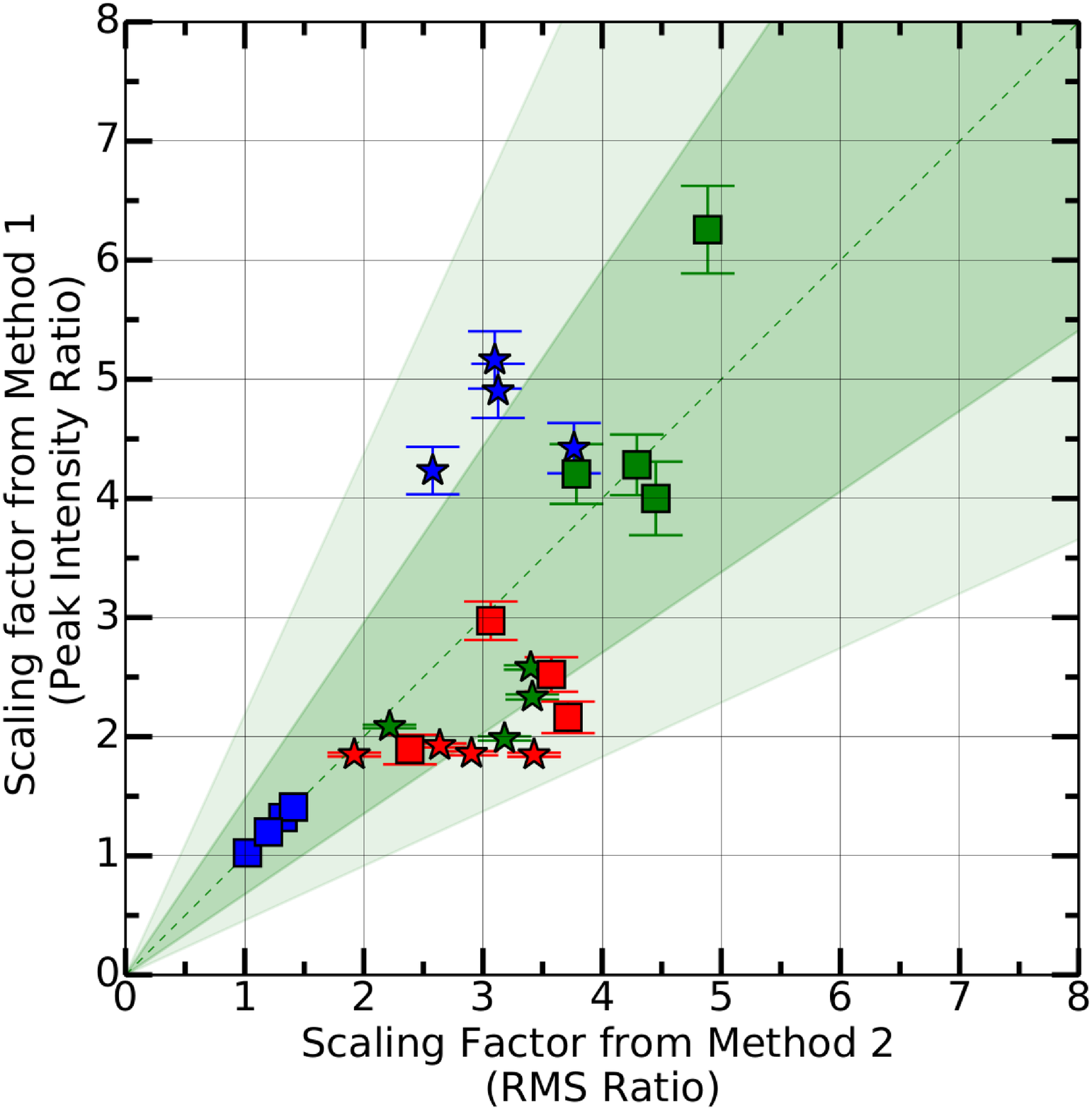}
    
    \caption{Scaling factors from Method 1 vs.\ Method 2 for observations on 03/25/2020. Darker shaded region is the $1\sigma$ dispersion (0.17~dex) while the lighter shaded region is the $2\sigma$ dispersion. Red stars are for J1206+3941. Green stars are for J0956+5753. Blue stars are for J0941-1335. Blue squares are for NGC 2992, red squares are for NGC 4151, and green squares are for NGC 3079.}
  \label{fig: Method Ratio D}
\end{figure}

\begin{figure}[h]
    \includegraphics[width=\columnwidth]{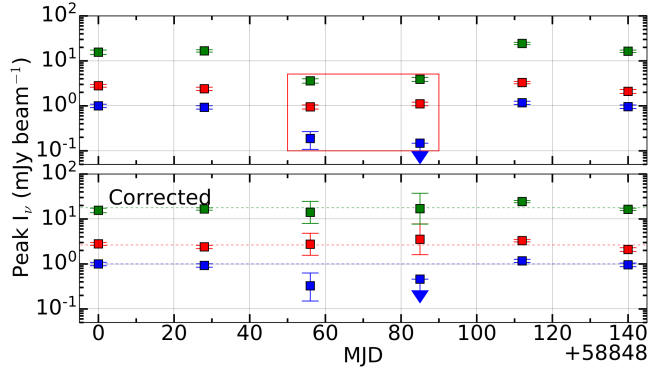}
    
    \caption{Source's peak intensity. Top: Before Corrections. Bottom: After corrections using their respective scaling factors. Uncertainties for affected observations calculated using Monte Carlo to obtain 95.4\% confidence intervals. \edit1{Blue squares are for NGC 2992, Red squares are for NGC 4151, and green squares are for NGC 3079. Dotted lines are their respective weighted averages using the non-affected observations.} Enclosed red box: Delineation caused by systematic issues. }
  \label{fig: Sources}
\end{figure}

Since the scaling factors obtained by the two methods are independent of each other, we first compare methods 1 and 2 by plotting the individual scaling factors as shown in Figures \ref{fig: Method Ratio C} and \ref{fig: Method Ratio D}. We also plotted the scaling factors for NGC~2992 based on the RMS method and inferred their location in the plot, which was not included in the scaling factor analysis, based on its RMS values.

There is an overall linear relationship between both methods, indicating that the scaling factor is different for each object. Since the scaling factor is multiplicative, we examine the reduced chi-squared $\chi^{2}_\mathrm{red}$ in log space to determine the difference between the two methods. The calculated $\chi^{2}_\mathrm{red}$ is 72.8 for 02/25/2020 and 60.7 for 03/25/2020, indicating additional uncertainty $s$, which we \edit1{calculated} by solving for $\chi^{2}_\mathrm{red}=1$: 

\begin{equation} \label{eq: Reduced Chi2 dispersion}
        1 = \frac{1}{N-1} \sum_{i=1}^{N}\frac{(\mathrm{MD}_i - \mathrm{\bar{MD}})^2}{\sigma_{\mathrm{MD}_i}^{2} + \mathrm{s}^{2}}
\end{equation}

\noindent MD are the logarithmic differences \edit1{between} methods 1 and 2. We find that $s=0.12$~dex and 0.17~dex for 02/25/2020 and 03/25/2020 respectively. Since Method 2 uses the most data and is unaffected by any real intrinsic variability as may be seen in AGNs, we use Method 2 to estimate the scaling factor, and add in quadrature the additional uncertainty $s$. To reiterate: both methods 1 and 2 compare the two affected observation dates with the four unaffected observation dates, so we take as the fiducial scaling factor for each affected data the average ratio between the RMS of the observation with the four values from the unaffected data. As only the formal flux uncertainties were included in the error term $\sigma_{\mathrm{MD}}$, the intrinsic uncertainty term $s$ gives the dispersion on the mean per-object scaling factor.

To robustly estimate the errors on the flux densities (peak and integrated), we performed a Monte Carlo simulation for each object, with $10^6$ random draws. On each draw, a random value from a log-normal distribution with sigma equal to the intrinsic dispersion was taken \edit1{(the intrinsic dispersion is multiplicative and therefore calculated in log space)}, and converted to a linear value. To this we added a random value from a normal distribution with sigma equal to the formal flux density uncertainty. The final error was associated to the flux density and stored in a vector, from which we derived the final 95.4\% confidence interval.

For NGC 2992, we take the conservative approach and use the largest scaling factor for each affected data set, 1.7 for the February data and 1.4 for the March data. Each object's scaling factor is used to correct their respective RMS and intensities of the affected observations. These corrections are included in the bottom panels of Figures \ref{fig: Phase_Cal}, \ref{fig: DB Phase Cal}, \ref{fig: DB Sources}, and \ref{fig: Sources}, and in \ref{fig: xray-radio} and also included in table \ref{tab:vlbadata}. 

One \edit1{caveat} to note: NGC 2992's RMS for the affected observations do not follow the same trend when compared to the phase calibrators and sources. The RMS values are consistent throughout the 6 month survey. The only factor that appears to be different is NGC 2992's RMS are significantly closer to the theoretical RMS. The ratios of the RMS and theoretical RMS for the other sources of the affected observations are all $>6$ while NGC 2992 is $\sim$2. 
To err on the side of caution, however, we apply a scaling factor to the NGC~2992 data, which has the effect of \emph{lessening} the observed radio variability. 

Figure \ref{fig: Sources} shows the peak fluxes \edit1{which includes} the two other FRAMEx sources, NGC 4151 and NGC 3079, before and after applying our scaling correction to the problematic datasets. From this analysis we see that the respective scale factors when applied to the other targets in our sample do a robust job in correcting the fluxes during the trouble epochs. We note that NGC 4151 is only $\sim2$ times the average flux of NGC 2992 implying that our non-detections during the sessions that suffered the software bug problems are, in fact, likely real. Through private communications with VLBA experts at NRAO, we employed numerous methods both in calibration and imaging to correct these troubled sessions but in the end, we find the scaling method described above most accurately and appropriately corrected our amplitudes.

\subsection{\textit{Swift}-XRT Observation} \label{subsection: SwiftXRt}

We obtained observation time with Target of Opportunity (ToO) (PI:~N.\ Secrest) using the Swift XRT, which has a PSF with half power diameter of $18\arcsec$ at 1.5~keV and a positional accuracy of $3\arcsec$. We requested integration time of 1.8~ks using Photon Counting (PC) mode and generated the X-ray spectra using the online XRT product generator \citep{2009MNRAS.397.1177E}, setting the same coordinates as the VLBA targeting coordinates. During our requested observation time in February 2020 and April 2020, other science projects took priority. Therefore, we only have simultaneous X-ray data for 4 out of 6 VLBA observations.

\subsubsection{X-ray Analysis}

\begin{figure*}
  \includegraphics[angle=-90,width =\textwidth]{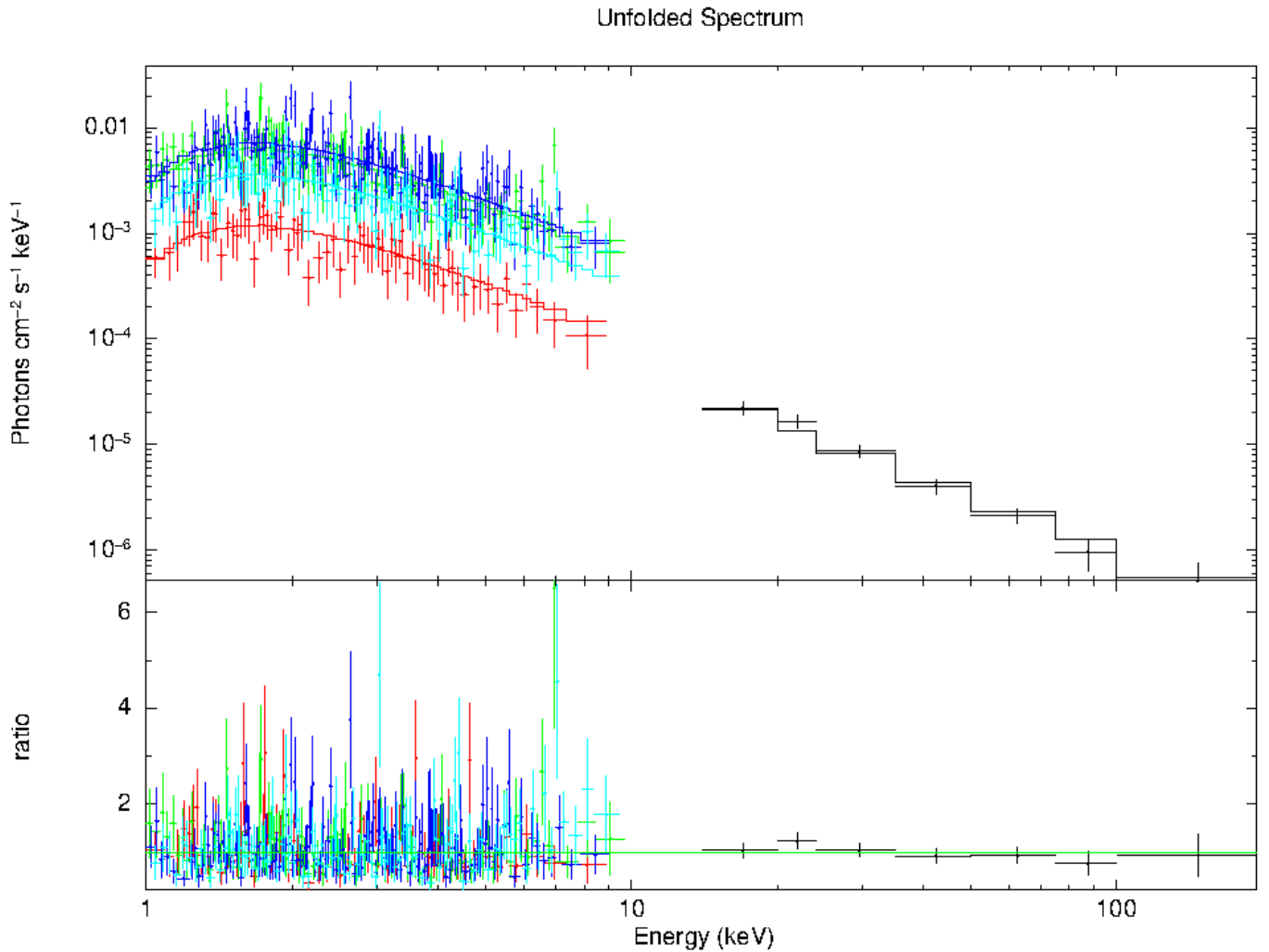}
    \caption{NGC 2992 X-ray spectrum fitting using the phenomenological power law model. 105-month BAT survey (Black). XRT observations: 12/31/2019 (Red), 01/28/2020 (Green), 03/25/2020 (Blue) , 05/19/2020 (Cyan).}
  \label{fig: xrayspec}
\end{figure*}

Spectral analysis was performed using \textsc{xspec} \texttt{v.12.11.1} \citep{1996ASPC..101...17A} software. We check for variability using our XRT data alongside the 8-band spectrum from the 105-month BAT catalog. This spectrum covers the period between December 2004 and August 2013 and has an effective integration time of 5.9~yr, providing a measure of the average intrinsic hard X-ray luminosity of NGC~2992. Folding in both XRT and BAT data spectrum then provides the best way to see how our short-term data varies from the long-term average. We use a simple absorbed power-law model (\texttt{phabs*zphabs*zpow}) to fit the X-ray spectra (shown in Figure \ref{fig: xrayspec}). There is soft excess seen in the 0.5 to 2~keV range as well as the possible appearance of Fe~K$\alpha$ and K$\beta$ lines in some of the observations. Fitting the spectrum with the physical model MYTorus \citep{2009MNRAS.397.1549M} helped to account for them, but did not produce an overall better fit, likely owing to a small column density of $N_\mathrm{H}\sim10^{22}$~cm$^{-2}$ \citep{2021ApJ...906...88F}. Therefore, we continue fitting the spectrum using the absorbed power law model and produce Monte Carlo Markov chains using the \texttt{chain} command to robustly estimate model errors and covariances.

Initially, we tied all parameters for the XRT data to the BAT data, but this resulted in a poor fit. We untied the power-law normalization, hydrogen column density, and photon index parameters in a number of combinations to see if there were any statistically significant differences between the free parameters. From our Markov chain analysis, we found that there is no significant evidence of variability in either the hydrogen column density $N_\mathrm{H}$ or the photon index $\Gamma$ when left as free parameters (or when one parameter is tied and the other is left free to vary). Only the variations of the normalization parameters were statistically significant, indicating that while the shape of the X-ray spectrum is invariant, the overall X-ray luminosity varies. Consequently, we tied $N_\mathrm{H}$ and $\Gamma$ for all datasets, and ran Markov chains to calculate statistical uncertainty on the intrinsic 2--10 keV flux.

\begin{deluxetable*}{c c c c}

\tablehead{\colhead{Observation Date} & \colhead{Normalization$^a$}&                               \colhead{$\log_{10}(F_\mathrm{2-10~keV}^a$ / erg cm$^{-2}$ s$^{-1}$)} & \colhead{$\log_{10}(L_\mathrm{2-10~keV}$ / erg  s$^{-1}$)}\\ [-0.1cm]
           & (${\times}10^{-2}$) &  & 
           }
\startdata
12/31/20 &  $0.61_{-0.12}^{+0.16}$ &  $-10.65_{-0.05}^{+0.04}$ & 42.47$_{-0.05}^{+0.04}$\\
01/28/20 &  $3.31_{-0.61}^{+0.85}$ &  $-09.92_{-0.04}^{+0.04}$ & 43.20$_{-0.04}^{+0.04}$\\
03/25/20 &  $3.71_{-0.67}^{+0.86}$ &  $-09.87_{-0.03}^{+0.03}$ & 43.25$_{-0.03}^{+0.03}$\\
05/19/20 &  $1.84_{-0.30}^{+0.46}$ &  $-10.18_{-0.04}^{+0.04}$ & 42.94$_{-0.04}^{+0.04}$\\
\enddata
\label{tab:XRTdata}
$^a$All uncertainties calculated are 95.4\% confidence interval
\caption{X-ray spectral fitting results of NGC 2992. Using the phenomenological model \texttt{phabs*zphabs*zpow} to fit the contemporaneous data, only the normalization have changes that are statistically significant. Uncertainties for the normalization and flux is calculated using Markov chains. The rest of the parameters are tied to the 105-month BAT survey giving a total statistic/dof of 1184.74/1421, an N$_H$ of $1.0_{-0.1}^{+0.2}\times10^{22}$ cm$^{-2}$ and a photon index ($\Gamma$) of $1.8_{-0.1}^{+0.2}$.}
\centering
\end{deluxetable*}

\section{Results} \label{section: results}

\subsection{Flux Variability at 6 GHz}
Examining the the results from the calibrated and imaged data, NGC 2992 was \edit1{marginally detected on 02/25/20 and }not detected in our observation on 03/25/2020. It appears to drop below our detection limit given our integration time of $\sim$1 hour. Using the scaling factors (from section \ref{subsubsection: Analysis}), we calculate the 3$\sigma$ upper limit to be \edit1{0.5 mJy} for this observation. The \edit1{peak flux density falls to less than half the average of $\sim$1 mJy.}

This type of variability has not been typically seen in radio quiet AGNs, largely because there are almost no studies on radio variability in these types of objects and even fewer at the resolution of the VLBA (let alone simultaneous X-ray and radio studies to probe the inner workings of radio quiet AGNs). Of those found in the literature on radio quiet AGNs, \cite{2001ASPC..224..265F} reports on two surveys; radio quiet quasars (RQQs) observed with the VLA and low luminosity AGNs (LLAGNs) observed with the VLA and VLBA. For both the RQQs and LLAGNs, a large number were found to be variable over the span of a year. \cite{2009ApJ...703..802M} examined Seyfert galaxies with the VLA and found flux variations over a 7 year period. Out of 12 detected sources 5 where found to be variable, excluding NGC 2110 which is a radio loud AGN, the average variation is $\sim0.3$~mJy, or $\sim45\%$. The only similar property found between them is compactness of their cores. The authors suggest that these sources might therefore exhibit variability. The remaining detected sources have an associated extended emission (either jet-like or non-relativistic). Only NGC 2110 has a variable compact core and extended emission from a radio jet. Although both of these works have stated long-term radio monitoring campaigns are needed to better understand these objects, few exist in the literature.

Next, we examined the brightness temperature $T_b$ for the peak intensity values found in table \ref{tab:vlbadata}. Starting with the Rayleigh-Jeans limit for brightness temperature \citep{2016era..book.....C}, we converted this expression by substituting the flux density to brightness units per solid angle where the solid angle is converted to the area of a Gaussian beam. All of the constants were combined into one term with units (including K) that cancel the remaining variables leaving only unit K. The brightness temperature expression for radio emission is therefore expressed as  

\begin{equation}\label{eq: tb_NRAO}
    T_b = 1.222 \times 10^3 \frac{I_\nu}{\nu^{2}\theta_{max}\theta_{min}}
\end{equation}

\noindent where \textit{I$_\nu$} is the peak intensity in units mJy/Beam, $\nu$ is in units GHz, $\theta_{max}$ and $\theta_{min}$ are the Gaussian major and minor axis half-power beam widths, in units arc-seconds, used to determine the peak flux. The corresponding brightness temperature is $\sim10^{6}$~K, but because the source is unresolved, this is a hard lower limit.

The presence of the unresolved radio variability allows us to set better constraints on the minimum brightness temperature $T_b$, which can yield insights into the size and nature of the emitting region. For a spherical black body, the Rayleigh-Jeans expression for the luminosity density is:

\begin{equation} \label{eq: rayleighjeans}
L_\nu = \frac{2 \nu^{2} k_B T_b}{c^2} 4 \pi r^2,
\end{equation}

\noindent where $r$ is the radius of the emitting source. Given variability in $L_\nu$ with some characteristic timescale $\tau$, the size of the source is set by $r = c \tau$. Rearranging Equation~\ref{eq: rayleighjeans}, the brightness temperature is:

\begin{equation} \label{eq: tb}
T_b = \frac{L_\nu}{8 \pi \nu^2 k_B \tau^2},
\end{equation}
\noindent which is equivalent to Equation~6 in \citet{2015ApJ...806..224M} in the limit of $v/c \ll 1$ (material not expanding at a significant fraction of the speed of light, such as a jet). There could, however, be variations in the radio on shorter timescales than what our observations sample, so Equation~\ref{eq: tb} provides a minimum brightness temperature of the emitting source. For a timescale of 28~days corresponding to the sampling of our observations and the minimum observed luminosity from Table~\ref{tab:vlbadata}, $T_{b,\mathrm{min}}\sim10^9$~K. This brightness temperature, combined with the overall low radio luminosity of the AGN, strongly favors a self-absorbed synchrotron source consistent with the hot compact hard X-ray corona.

\subsection{Soft X-ray Variability}
We show the X-ray spectra of NGC~2992 in Figure~\ref{fig: xrayspec}. Using a simple power law model, we find that there is variability consistent with what has been found previously \citep{2007ApJ...666...96M,2020MNRAS.tmp.1826M} in the 2--10\,keV regime. The photon index ($\Gamma$) and column density ($N_\mathrm{H}$) does not vary from epoch to epoch, with $N_\mathrm{H}=1.0_{-0.1}^{+0.2}\times10^{22}$ cm$^{-2}$ and $\Gamma=1.8_{-0.1}^{+0.2}$. Table~\ref{tab:XRTdata} contains the power-law normalization variations from the fitted data with the calculated logarithmic $F_\mathrm{2-10~keV}$ and $L_\mathrm{2-10~keV}$.

\subsection{X-ray and Radio Anti-Correlated}

\begin{figure}[h]

  \includegraphics[width=\columnwidth]{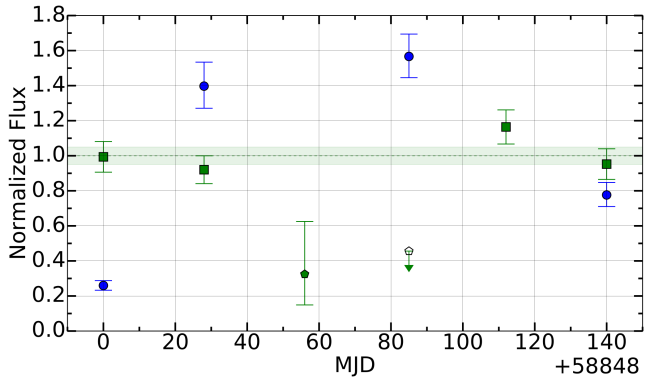}
    \caption{Normalized radio and X-ray flux (all uncertainties are $\pm2\sigma$). Green squares denote normalized C-band (6 GHz) radio flux (normalized by 5.72$\times$10$^{-17}$ erg cm$^{-2}$ s$^{-1}$). Blue circles denote normalized 2-10 keV X-ray flux (normalized by 8.60$\times$10$^{-11}$ erg cm$^{-2}$ s$^{-1}$). Green pentagon denotes radio observations corrected using their respective scaling factors with the uncertainty calculated using Monte Carlo to obtain 95.4\% confidence interval. The upper limit (3$\sigma$ over the RMS) denoted by downward arrow. Green dotted horizontal line is concatenated VLBA data from all non-affected observations with uncertainty.}
  \label{fig: xray-radio}

\end{figure}

In Figure \ref{fig: xray-radio} we show the normalized flux for the contemporaneous X-ray (2--10 keV) and radio ($\sim$6 GHz) observations. There appears to be an anti-correlation in which when the X-ray flares, the radio emission is attenuated and as the X-ray flux diminishes, the radio emission re-appears. While we lost simultaneous observations in February and April, the overall anti-correlation trend is nonetheless clear.

\subsection{Faint Secondary Radio Emission}

Examining the image from the concatenated non-affected data (Figure \ref{fig: Imaging Avg}), there appears to be evidence of a faint radio emission source to the SE of the core radio emission. It is separated by 22.3 mas from the peak intensity of the core radio emission to that of the faint source. Located at RA 146\fdg4247702 and DEC $-$14\fdg3262833. It has a significance of $\geq$ 5$\sigma$. The faint emission has a peak intensity of 0.15 $\pm$ 0.02 mJy/beam which corresponds to a lower limit brightness temperature T$_b$ $\sim$10$^5$ K. This most closely resembles free-free emission seen similarly in NGC 1068 \citep{2004ApJ...613..794G}. It is unclear at this time if this is previously ejected material from the core of NGC 2992. Since this image is from multi-epoch observations, a follow-up observation is needed with longer integration time ($\sim$4 hours) to confirm the extent of the faint emission.

\section{Discussion} \label{section: discussion}

\subsection{Radio Variability}
There are a few physical mechanisms that can help explain the variability seen in our observations. We begin with the most likely scenario and discuss other possibilities.

\subsubsection{Free-Free Absorption}
The 6~GHz core radio emission in NGC~2992 exhibited a decline, by a factor of at least $\sim3$, over a 3-month period, in tandem with a $2-10$~keV flare, before recovering at the end of the flare. While the literature on the potential physical mechanisms behind simultaneous X-ray and radio variability of radio-quiet AGNs are lacking, some work on black hole binaries (BHBs) may be informative. For example, \citet{1999ApJ...519L.165F} report a drop in radio emission, which they report as a jet, in the BHB GX 339-4 throughout a period of high X-ray luminosity. Both the MOST (36 cm) and ATCA (3, 6, 20 cm) radio emission the BATSE $20-100$~keV flux plummet in tandem with an outburst in the $2-12$~keV RXTE ASM flux (see their Figure 1). The entire event occurs over a period of 400 days, and the behavior is attributed to the innermost accretion disk extending closer to the BH, diminishing the Comptonizing corona and extinguishing emission at hard X-rays. 

This ``high-soft, low-hard'' paradigm has been proposed to unify BHB and AGN accretion states \citep[e.g.,][]{2006MNRAS.372.1366K}; however, to date no evidence has been found for a similar anti-correlation between the hard and soft X-rays in NGC~2992, and the constancy of X-ray spectral index $\Gamma$ found here and in previous studies \citep[e.g.,][]{2007ApJ...666...96M, 2007ApJ...666..122B} argues strongly against this picture. Indeed, the constancy of the X-ray spectrum over a factor of $\sim8$ variation in apparent luminosity suggests, as was argued in \citet{2007ApJ...666..122B}, that the amount of Comptonizing plasma has varied, e.g.\ as might be caused by magnetic reconnection events in the innermost accretion disk \citep[e.g.,][]{1999MNRAS.306L..31P, 2010A&A...518A...5D}. In this case, a burst of Comptonizing plasma increases the bulk number of up-scattered UV photons from the accretion disk, leading to a change in X-ray luminosity without a significant variation in spectral index.

A possible explanation for the simultaneous radio variability of NGC~2992 is changes in free-free absorption. Given that the decrease in radio emission is accompanied by an increase in the X-ray flux, this indicates higher levels of activity, which can result in ionized material being ejected from the accretion disk, either a cloud or a wind, and potentially pass between of the nuclear radio source and the observer, increasing the opacity at radio frequencies and dimming the flux. 

We have from Osterbrock (1989), that the optical depth due to free-free absorption can be calculated using the following expression: 
     
\begin{equation} \label{eq: freefree}
\tau_{\nu} = 3.28 \times 10^{-7} \left(\frac{T}{10^4~\mathrm{K}}\right)^{-1.35} \left(\frac{\nu}{\mathrm{GHz}}\right)^{-2.1} E
\end{equation}

\noindent where $T$ is the temperature of the ionized gas in units of $10^4$~K, $\nu$ is the observed frequency in GHz, and $E$ is the emission measure, which corresponds to the following expression:

\begin{equation} \label{eq: emission measure}
E = \left(\frac{ \int n_+ n_e \,ds }{\mathrm{pc~cm^{-6}}}\right)
\end{equation}

\noindent We can assume that $n_+ \sim n_e$, so the emission measure is the integral of $n_e^2$ along the line of sight. Based on the values from Table~\ref{tab:vlbadata} we can calculate that, relative to the average flux of the first 2 and last 2 data points, one would need an ionized sources with an optical depth $\tau_\mathrm{6~GHz}=\edit1{0.71}$\footnote{Calculations used exact frequency of 5.8~GHz.}, in order to reduce the 2020-March-25 flux to the observed  $3\sigma$ upper limit of \edit1{0.5~mJy}.

\begin{figure}[h]
  \includegraphics[width=\columnwidth]{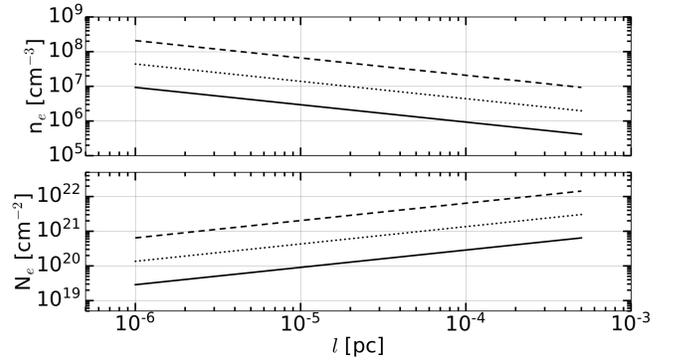}
    \caption{The top panel shows the electron density ($n_e$) as a function of cloud thickness ($l$), that would result in $\tau_{6GHz}=\edit1{0.71}$, for ionized gas with temperature of 10$^4$, 10$^5$ and 10$^6$ K (solid, dotted and dashed lines respectively). The bottom panel shows the corresponding electron column densities (N$_e$).}
  \label{fig:freefree}
\end{figure}

\begin{figure*}[h]
  \includegraphics[width = \textwidth]{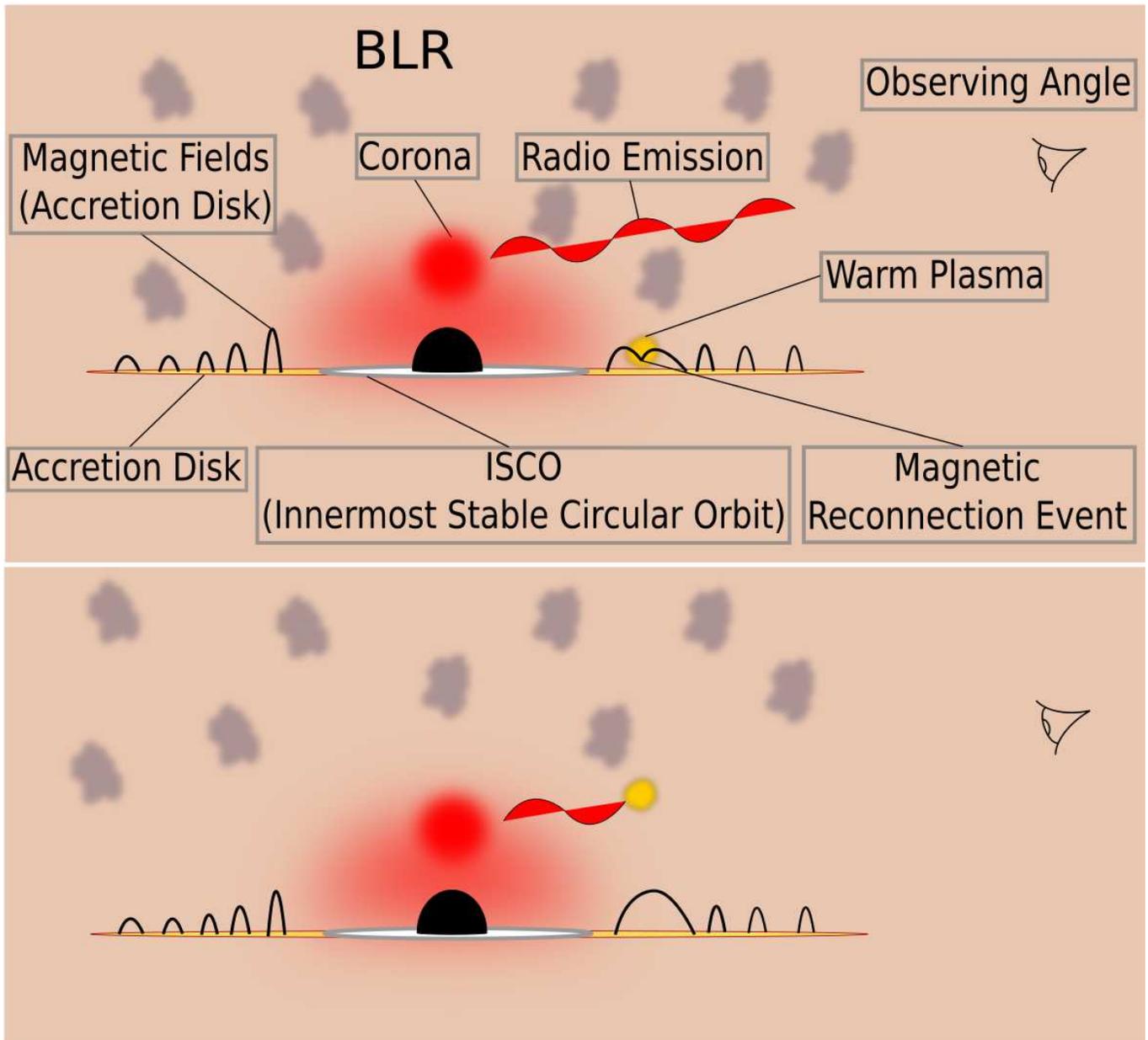}
    \caption{Top: Warm clumpy plasma cloud above accretion disk before being ejected due to magnetic re-connection events into the broad line region (BLR). This reflects the observations before the drop in intensity. Bottom: Plasma cloud causing free-free absorption affecting the radio luminosity obscuring our line of sight. This reflects the observation on 03/25/2020.}
  \label{fig: Qualitative Model}
\end{figure*}

Combining this optical depth value with Eq.~\ref{eq: freefree}, we were able to calculate the electron density ($n_e$) as a function of thickness of the intervening ionized region, for temperatures $T=10^4$, $10^5$, and $10^6$~K. We also calculated the corresponding electron column densities ($N_e$). These results are presented in Fig.~\ref{fig:freefree}, where we can see that an intervening cloud with a thickness in the range $10^{-6}\leq l \leq 5\times10^{-4}$~pc and electron densities in the range $\edit1{4}\times10^5\leq n_e \leq \edit1{2}\times10^8$~cm$^{-3}$ will result in this optical depth. These values correspond to typical conditions in the broad line region, as well as in intermediate regions between the broad and the narrow line regions. We also find that these values correspond to electron column densities in the range $\edit1{3}\times10^{19} \leq N_e \leq \edit1{1}\times10^{22}$~cm$^{-2}$, which are consistent with the fact that the X-ray observations do not show a significant change in column density. In Figure \ref{fig: Qualitative Model} we show a qualitative model for the free-free absorber. This begins with magnetic re-connection events that launch clumpy dense warm plasma causing the drop in intensity seen in our radio observations.  

This phenomena is similarly seen in X-ray binary systems where recently, \cite{2021MNRAS.500.4166S} looked at X-ray binary system LS I $+61^{\circ}303$, and found it to be variable in both X-ray (0.3 - 10 keV) and radio (13–15.5 \& 15.5–18 GHz) with significant correlation . They suggest both emissions are due to the same electron population, but it is unknown if both emissions are caused by a singular physical process or if multiple processes contribute individually to each emission. This was observed in an optically thin flare which is most likely due to shocks or magnetic re-connection events. Given multiple re-connection events, this can lead to multiple ejection of plasmoids of different sizes. The frequency of these events can be anywhere from minutes, hours, and days. They can take place in the accretion disk to produce flares which have been observed in hard to soft state transitions in microquasars (seen in GRS 1915+105). These flares eject plasmoids as single blobs that can expand almost adiabatically which can quickly become optically thin in the radio band after leaving the accretion disc \citep[e.g.][]{2009MNRAS.395.2183Y}, although the time frame for this process is not clearly stated.

\subsubsection{Intrinsic Variability}
Considering the variability transpires within 28 days, this confines the emitting region for both the (2-10 keV) X-ray and (6 GHz) radio luminosity (with X-ray variability seen in previous studies on shorter time scales of days to weeks) to within a radius of 28 light days. With the resolution of the VLBA and at these time scales, the physical mechanism for the luminosity changes must be confined to within this radius and not from any larger structures crossing the line of sight. This leads to the possibility the variability in the radio and X-ray emission are intrinsic for NGC 2992 and stem from the same electron population in the corona. This may explain why during the flare when the X-ray peaks, the radio drops below our detection limit. However, this does not explain why the radio is delayed at the beginning of the flare. In the literature there have been delays observed, but it is when the radio preceded the X-ray. In one such extreme case seen in blazar PKS 1510-089, radio emission precedes the X-ray emission by 24 days \citep[e.g.][]{2010ApJ...710L.126M}. For the X-ray binary system LS I $+61^{\circ}303$, the radio preceded the X-ray by $\sim$25 minutes \citep{2021MNRAS.500.4166S}. Since our observations are in 28 day intervals, we do not have enough data to determine if the variability seen in the radio varies at the same time scales as seen previously with X-ray (days to weeks). It is possible the true delay is in fact the radio preceding the X-ray and not the X-ray leading the radio as shown from our observations. Finally, there is also the possibility the radio and X-ray variability are unrelated and have separate physical mechanisms. Future simultaneous observations are needed in shorter intervals (weekly to twice a week) to determine which follows for NGC 2992.

\section{Conclusions} \label{section: conclusions}

To our knowledge, this is the first simultaneous X-ray and VLBI radio monitoring campaign of a nearby radio-quiet AGN. Our main conclusions are as follows:

\begin{enumerate}

    \item We find anti-correlated core radio (6~GHz) and X-ray (2--10~keV) emission from the known X-ray variable AGN in NGC~2992. The radio emission declines by over a factor of $>3$, shortly after (within 28~days of) a flare in the 2--10~keV X-ray emission (by a factor of $\sim6$. The size of the radio-emitting region is constrained by the variability, and is consistent with radio emission originating within the central accretion region ($<0.02$).
     
    \item Given the current understanding of AGN accretion, the simultaneous X-ray and core radio behavior seen in NGC~2992 can most naturally be understood as being due to flares produced by magnetic reconnection events in the accretion disk. These flares create outbursts of Comptonizing plasma, leading to an overall brightening in hard X-rays, and some of the material enters the broad line region, increasing free-free absorption at radio wavelengths.\\
    
\end{enumerate}

We have taken care to robustly estimate the influence of the two NGC~2992 observations potentially affected by the software bug in the new Mark~6 recorders, and we included a scaling factor to correct for this issue. This scaling factor included a robustly-estimated intrinsic uncertainty term that we have included in our analysis.
Nonetheless, including the scaling factor and its uncertainty is the more conservative approach, and so we are confident that the radio variability we have observed in NGC~2992 is real.

\acknowledgements

This research made use of Astropy,\footnote{\url{http://www.astropy.org}} a community-developed core Python package for Astronomy \citep{2013A&A...558A..33A, 2018AJ....156..123A}, as well as \textsc{topcat} \citep{2005ASPC..347...29T}. L.C.F.\ is thankful for the support from his spouse Kathryn Fern\'andez. 

The National Radio Astronomy Observatory is a facility of the National Science Foundation operated under cooperative agreement by Associated Universities, Inc. The authors acknowledge use of the Very Long Baseline Array under the U.S.\ Naval Observatory’s time allocation. This work supports USNO’s ongoing research into the celestial reference frame and geodesy.

\facilities{VLBA, SWIFT}

\software{
Astropy \citep{2013A&A...558A..33A,2018AJ....156..123A}, 
\textsc{TOPCAT} \citep{2005ASPC..347...29T},
\textsc{AIPS} \citep{2003ASSL..285..109G},
\textsc{XSPEC} \citep{1996ASPC..101...17A}}

\bibliography{Fernandez_et_al_2022}{}
\bibliographystyle{aasjournal}

\end{document}